\documentstyle[draft,epsf,psfig]{mn}

\def\gs{\mathrel{\raise0.35ex\hbox{$\scriptstyle >$}\kern-0.6em 
\lower0.40ex\hbox{{$\scriptstyle \sim$}}}}
\def\ls{\mathrel{\raise0.35ex\hbox{$\scriptstyle <$}\kern-0.6em 
\lower0.40ex\hbox{{$\scriptstyle \sim$}}}}

\def\Gyr{\,\hbox{Gyr}}

\title[The Colour-Magnitude Relation]
	{The Colour-Magnitude Relation as a 
	Constraint on the Formation of Rich Cluster Galaxies}
\author[Bower, Kodama \& Terlevich]
{
 Richard G. Bower$^1$, Tadayuki Kodama$^2$ \& Ale Terlevich$^1$\\
	$^1$ Department of Physics, University of Durham, South Road., 
		Durham DH1 3LE, UK\\
	$^2$ Institute of Astronomy, University of Cambridge, Madingley Road, 
		Cambridge CB3 0HA, UK
}

\begin{document}

\label{firstpage}

\maketitle

\begin{abstract}

The colours and magnitudes of early-type galaxies in galaxy clusters
are strongly correlated. The existence of such a correlation has been
used to infer that early-type galaxies must be old passively evolving 
systems. Given the dominance of early-type galaxies in the cores of rich 
clusters, this view sits uncomfortably with the increasing fraction of 
blue galaxies found in clusters at intermediate redshifts, and with the 
late formation of galaxies favoured by CDM-type cosmologies. 
In this paper, we make a detailed investigation of these issues and 
examine the role that the colour-magnitude relation can play in constraining
the formation history of galaxies currently found in the cores of rich 
clusters. We start by considering the colour evolution of galaxies after 
star formation ceases. We show that the scatter of the colour-magnitude 
relation places a strong constraint on the spread in age that is allowed 
for the bulk of the stellar population. In the extreme case that the stars 
are formed in a single event, the spread in age 
cannot be more than 4 Gyr. Although the bulk
of stars must be formed in a short period, continuing formation of stars
in a fraction of the galaxies is not so strongly constrained. We examine 
a model in which star formation occurs over an extended period of time
in most galaxies with star formation being truncated randomly. This
model is consistent with the formation of stars in a few systems until
look-back times of $\sim 5\Gyr$. An extension of this type of star formation
history allows us to reconcile the small present-day scatter of the 
colour-magnitude relation with the observed blue galaxy fractions of
intermediate redshift galaxy clusters.

In addition to setting a limit on the variations in luminosity-weighted
age between the stellar populations of cluster galaxies, the 
colour-magnitude relation can also be used to constrain the degree of merging 
between pre-existing stellar systems. This test relies on the slope of 
the colour-magnitude relation: mergers between galaxies of unequal mass 
tend to reduce the slope of the relation and to increase its
scatter. We show that random 
mergers between galaxies very rapidly removes any well-defined 
colour-magnitude correlation. This model is not physically motivated,
however, and we prefer to examine the merger process using a self-consistent 
merger tree. In such a model there are two effects. Firstly, massive galaxies 
preferentially merge with systems of similar mass. Secondly, the rate
of mass growth is considerably smaller than for the random merger case.
As a result of both of these effects, the colour-magnitude correlation 
persists through a larger number of merger steps. 

The passive evolution of galaxy colours and their averaging 
in dissipationless mergers provide opposing constraints on the formation
of cluster galaxies in a hierarchical model. At the level of current 
constraints, a compromise solution appears possible. The bulk of the stellar
population must have formed before $z=1$, but cannot have formed in mass units
much less than about half the mass of a present-day $L_*$ galaxy.  
In this case, the galaxies are on average old enough that stellar population 
evolution is weak, yet formed recently enough that mass growth due to
mergers is small.

\end{abstract}


\section{Introduction}

The existence of a strong correlation between the colour and luminosity
of early-type galaxies in rich clusters is well established 
(eg., Sandage \& Visvanathan, 1978, Larson, Tinsley \& Caldwell, 1980).
This colour-magnitude relation (CMR) implies a physical link 
between the stellar populations of galaxies and their
total stellar mass. The best explanation appears to be a correlation
between the metal abundance of the stellar population and
total galaxy mass (eg., Arimoto \& Yoshii, 1987, Bower, Lucey \& Ellis, 
1992b, BLE), although explanations based on the age of galaxies and on 
their stellar initial mass function (IMF) have also been proposed 
(eg., Trager et al., 1993, Tantalo et al., 1997). Metal abundance variations 
are widely accepted as the more likely explanation because observations of 
the CMR in high redshift clusters show a uniform shift towards bluer 
colours for galaxies of all magnitudes while maintaining small scatter about 
the mean relation (Ellis et al., 1997, Kodama \& Arimoto 1997, 
Stanford, Eisenhardt \& Dickinson 1997, Kodama et al.\ 1998).  
As we discuss below, the tightness of the CMR leaves little room for 
additional variations in galaxy age.

In this paper, we explore the constraints that can be placed on the 
formation of galaxies by the existence of the CMR. Through out this
paper, we will assume that the metal abundance and age of a galaxy are
uncorrelated --- this is the case for both the super-novae wind models 
of Arimoto \& Yoshii (1987) and for self-regulated hierarchical
galaxy formation models (White \& Frenk, 1991, Kauffmann et al., 1993, 
Cole et al., 1994, and subsequent papers). 
Ferras et al.\ (1998) provide an extended discussion of the formal
limits on correlated variations.
We focus on the scatter of galaxies about the mean correlation and on
the slope of the correlation relative to the observed scatter.
For galaxies in clusters the rms residuals are remarkably small, less than 
0.05~mag rms in $U-V$ for E and S0 galaxies in the Coma cluster. This implies 
a marked homogeneity in the galaxies' formation that provides an important 
constraint for theoretical models.
For example, if the general trend of increasingly blue colours
at fainter magnitudes were due to age, the galaxies would be required 
to range in `age' (ie., the look-back time to the epoch at which the 
bulk of the stars were formed) by $\sim 7$ Gyr across the first 3 mag 
of the relation (with the larger galaxies being formed at earlier times), 
while being coordinated so that galaxies of the same magnitude 
`formed' with an rms dispersion substantially less than this. Other models 
for the origin of the relation are similarly constrained.  Under the 
assumption that the CMR is driven by age, BLE used this argument 
to suggest that most galaxies must have formed the bulk of their stars at 
redshifts greater than unity. 
The CMR data also limit the possible contribution of bursts of star formation
after the formation of the main stellar component and the degree to which 
present-day galaxies can have been built-up by the merger of stellar (as 
opposed to gaseous) systems.

Taken in isolation, the tightness of the colour-magnitude relation
suggests that most galaxies in clusters are old systems, but
this picture appears to conflict with observations of intermediate redshift
clusters. Photometric studies (eg. Butcher \& Oemler, 1978, 1984, Couch 
\& Newell, 1984) have suggested very significant changes to the clusters' 
mix of galaxy types over look-back times of 3-5 Gyr. Probably, this change 
is associated with the transformation of Spiral galaxies into S0 types
(Dressler et al., 1997, Couch et al., 1997). This picture is reinforced 
by spectroscopic observations showing that a large fraction of the galaxies 
in these clusters have only just completed their star formation cycle 
(eg., Dressler \& Gunn, 1983, Couch \& Sharples, 1987, Barger et al., 1996). 
Can this evidence of star formation activity at relatively modest 
look-back times be compatible with the homogeneity of the present-day galaxy 
populations? To answer this question, we must carefully examine the 
scatter that would be introduced by a population of galaxies that were 
forming stars in the recent past. In BLE's data for the Coma cluster,
for example, a small number of galaxies do not fit on the ridge-line of 
the CMR. It is important that we correctly account for the
treatment of out-lying galaxies when comparing the present-day and
intermediate redshift observations, and that we allow for the full range
of morphological types seen in present-day clusters.

The well-defined slope of the CMR implies that the star 
formation process is linked to the final mass of the system: if most
of the star formation were to occur in systems of low mass, a physical
link between the initial systems and their final mass would be required.
In the absence of such a process, the CMR sets a strong constraint on the 
number of pieces that can be combined to form the final system.
 
The approach outlined by BLE does not rely on a specific model for
galaxy formation in order to derive constraints from the observational
data. The short-coming of this approach is that it takes no account
of the inevitable gravitational evolution of the dark matter haloes
in which the galaxies are contained. For example, although the early-type
galaxies on which this data is based currently reside in rich clusters,
at $z=2$ this cluster scale over-density would not yet have 
collapsed; the typical dark matter halo been would have had a mass
close to that of a small galaxy group, and the finally galaxy would probably
not be recognisable as a single unit. White \& Frenk (1991),
Kauffmann et al.\ (1993), Cole et al.\ (1994) and subsequent papers, have 
therefore pursued an alternative approach.
They take as a starting point the extended Press-Schechter theory (Press \&
Schechter, 1974, Bower, 1991, Bond et al., 1991, Lacey \& Cole 1993). 
This allows Monte Carlo
simulation of the halo merger tree corresponding to a present-day halo
of a given mass. Simple prescriptions for the cooling of gas, formation of 
stars and the feedback of the energy released in supernovae, and the
merging and accretion of galaxies can then be fed into the realistic
gravitational framework. The rudimentary parameters controlling these
processes can then be adjusted until a realistic galaxy population
is created. Using these procedures, Kauffmann (1996) and Kauffmann \& Charlot
(1997, KC97) have argued that an elliptical galaxy colour-magnitude 
relation can be produced that accurately mimics the observed one, both in
its slope and in scatter about the mean relation. This occurs
despite the fact that star formation continues until relatively late
epochs in such hierarchical models (eg., 50\% of stars form at $z<1$ in the 
models of Baugh et al., 1997 [BCFL]), and that considerable merging occurs 
between stellar components. 

This paper re-examines the constraints which can be derived from the 
CMR in the light of these models. Following BLE, we limit 
the galaxy formation process in a model independent manner, 
while at the same time allowing for realistic growth of the gravitational 
potential. Our aim is to synthesize the model independent approach of BLE
with the all-inclusive modeling of KC97 and BCFL.
We consider a simplified scheme in which the bulk of star formation occurs
at relatively high redshift giving rise to a well defined CMR
for proto-galactic sub-units. Merging of these increases the mass
of the galaxy causing the initial relationship to evolve increasing
in scatter and decreasing in slope. Random merging of the sub-lumps
creates a relationship that is very flat and has considerable
scatter. The key issue we address is the extent to which merging
in a hierarchical universe leads to a predominance of equal mass mergers
and thus to a CMR that maintains it slope and keeps an
acceptably small scatter. To do this we use the galaxy merging trees from
BCFL but avoid using a detailed model for the galaxy formation process 
in order to isolate the effect of galaxy mergers. We do not attempt to 
assign morphological classifications to the model galaxies. Instead, we
compare the sample with a morphologically complete sample of galaxies in 
the core of the Coma cluster (Terlevich et al., 1998). 

In outline, the structure of this paper is as follows.
The first part of the paper is concerned with the conclusions that can be
drawn regarding the stellar populations of cluster galaxies.
In \S2, we revisit the scatter analysis of BLE. We consider a range 
of possible star formation histories 
ranging from continuous almost uniform star formation to star formation that 
declines strongly through the lifetime of the galaxy. We also consider
the effects of the clipping algorithm used to determine the scatter
from the observational data.
\S3 investigates whether the range of star formation histories considered
allow a consistent paradigm that explains both the narrowness of the 
CMR at low redshift and the evidence of star formation activity in 
intermediate redshift clusters. In \S4 we investigate the effects of mergers 
on the evolution of the CMR. In particular, we determine whether there is a 
conflict between the old ages implied for the bulk of the stars and 
the degree of merging implied by hierarchical gravitational evolution.

Finally, in \S5, we present a discussion of the results of our study. 
We show that the small scatter of
galaxies in the cores of rich clusters remains a challenging constraint
for realistic galaxy formation models. However, by weighting the 
rate at which star formation is truncated in the galaxies towards
large look-back times, and incorporation of the effects of the 
colour-magnitude fitting process, it is possible to explain both the 
increasing blue galaxy populations of intermediate clusters and the 
uniformity of the present-day galaxy populations. Furthermore, we show 
that while random merging quickly removes any initial slope of the
colour-magnitude relation, inclusion of the dynamical biases
inherent in a hierarchical model sustains the ratio of the scatter and slope 
of the colour-magnitude relation at a value close to that 
observed. A summary of our investigation is given in \S6.

\section{Constraining the history of star formation}

\subsection{Single-burst models}

In this section, we consider the constraints that 
the tightness of the colour-magnitude relation places on the star formation
histories of early-type galaxies. We base our investigation on the
$U-V$ colour as this colour straddles the 4000\AA\ break and has high 
sensitivity to evolution of the stellar population. We will adopt the
Coma cluster as the proto-type of rich present-day galaxy clusters.
High precision photometry in the $U$ and $V$ bands is available from
BLE and Terlevich et al., 1998. The arguments we put forward are generally 
applicable to all broad-band measures of stellar populations. 

We first consider the case in which early-type galaxy formation occurs
in a single short-lived burst. This type of formation history
might be suitable for the formation of elliptical galaxies (eg., Mathews
\& Baker, 1977, Arimoto \& Yoshi, 1987), although
such models over produce the numbers of very red galaxies
seen in deep number-count surveys (eg., Silk \& Zepf, 1996, Zepf, 1997)
and over-predict the numbers of morphologically classified E and S0
galaxies with $z>1$ in the Hubble Deep Field (Franceschini et al., 1998).
These problems can be avoided if an extensive component of dust is 
associated with the burst event. With an extensive dusty component, the 
proto-elliptical
galaxies would be luminous at infra-red wavelengths, and are therefore
open to detection by ISO (eg., Taniguchi et al. 1997) and sub-millimeter 
arrays such as SCUBA (cf., Smail et al., 1997). Such models are appealing,
however, since a strong correlation between the metal abundance of a 
galaxy and its mass arises naturally as a consequence of the more
massive galaxies greater binding energy.
\begin{figure}
\begin{center}
  \leavevmode
  \epsfxsize 1.0\hsize
  \epsffile{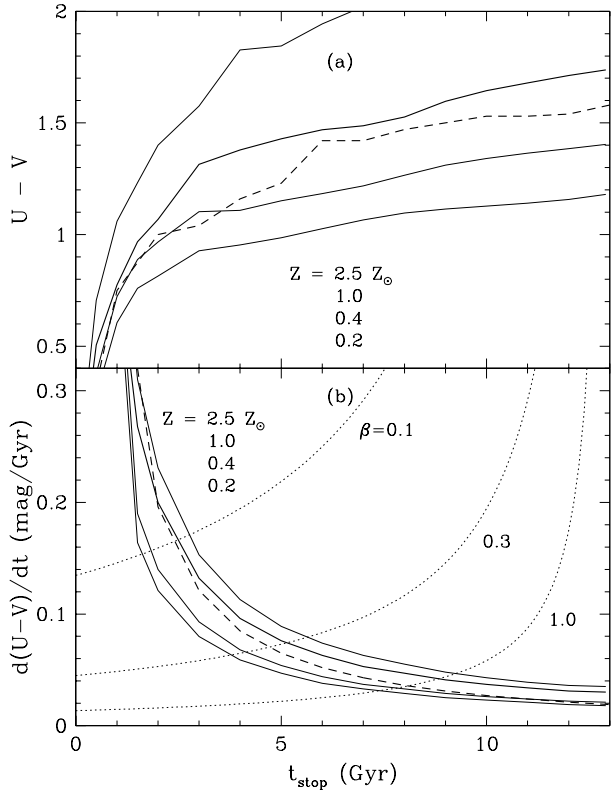}
\end{center}
\caption{(a) The U-V colour and (b) the rate of change of U-V colour 
(lower panel) as a function of the `age' of the stellar population.
In this case all stars are assumed to be formed in a single short-lived 
burst.
The curve shows the effect of varying the metal abundance of the stellar
population between $2.5 Z_\odot$ and $0.2 Z_\odot$. 
Note that a small degree smoothing applied to the curves in Figure (b).
While colour is sensitive to metal abundance, the rate of change of colour is 
almost the same over this very large range of values. For comparison,
the colour and its rate of change of a $Z_\odot$ stellar population
from Bruzual (1983) are plotted as dashed lines.
}
\label{fig:metratefig}
\end{figure}

The constraints on such burst models were considered extensively in 
BLE. Figure~\ref{fig:metratefig} shows the rate of colour
change as a function of the time since the star burst. This figure
is central to the analysis of BLE. The curve plotted uses the 
stellar population synthesis code of Kodama and Arimoto (1997)
with varying metal abundance. The figure also shows
the population synthesis code of Bruzual (1983): the rate of colour
change has altered little despite the substantial improvements
that have been made in the range of stellar populations that are
included in these codes. 
This reinforces the inherent simplicity of using this colour scheme
to study star formation history: changes in the $U-V$ colour are driven
primarily by the evolution of the main sequence turn-off, and are
thus relatively simple to model correctly.

This figure can be used to set limits to the spread in the formation
epochs of the early-type galaxies, and thus to infer the approximate
redshift range of galaxy formation. The rate of change of $U-V$ colour
gives the spread in colour of two galaxies formed a $\Delta t$
apart in age: $\Delta(U-V) \approx \Delta t. d(U-V)/dt$. Knowing
$\Delta(U-V)$ from our observational data and $d(U-V)/dt$ from our models
we can estimate $\Delta t$. Thus the
spread of galaxies about the mean colour-magnitude relation can
be converted into a spread in formation times. If we assume that 
galaxies form over a fraction $\beta$ of the time that is available,
this can in turn be converted in to an estimate of the time before which 
most galaxies were formed.

One simple approach is to assume that galaxy formation occurs with a uniform 
random distribution in look-back time between an early epoch, $t_0=13$~Gyr,
and a time $t_{stop}$ - a parameter that we wish to estimate. The value of
$t_0$ has been chosen to match the observed ages of the oldest
stars in our galaxy (eg., Pont et al., 1998, Gratton et al., 1998). 
In this case, the rms dispersion in galaxy colours is approximately 
$$
   \beta\left(t_0 - t_{stop}\over3.5\right).{d(U-V)\over dt}
$$
where the factor 3.5 relates the range in formation times to its
rms spread. Given the scatter in the observed CMR, and assuming a
value for $\beta$, this expression can be inverted to give a lower
limit on $t_{stop}$. The dotted curves shown in Figure~1 show the
limits that can be set from an observed scatter of 0.05~mag. We will
adopt this as a fiducial estimate in what follows.
In the case that galaxy formation occurs over the full range of 
available times, $\beta=1$. Lower values of $\beta$ either require galaxy 
formation to be more constrained (this is discussed below), or for the
observational scatter to be overestimated by BLE's approach (see \S2.2). 
Where the model $d(U-V)/dt$ lies above the fiducial $\beta=1$ line, the 
spread in galaxy formation
times results in too much scatter about the mean relation. This
forces $t_{stop}$ to lie above 9~Gyr for $\beta=1$ so that the age spread 
is less than 5~Gyr, and half of the galaxies are formed above 
$z\sim2$.\footnote{
	Throughout this paper, we will convert between age and redshift 
	assuming a cosmology with $q_0=0.5$ and $H_0 = 50\,\hbox{km/s/Mpc}$. 
	This gives an age of the universe of 13~Gyr. Note, however,
	that the arguments we present in \S2 are primarily dependent on the 
	cosmic look-back timescale and not on the redshift scale given by the
	assumed cosmology.} 
This constraint would be weakened if the galaxies are formed over a smaller 
time spread than is actual available (ie., $\beta<1$), but current 
evidence points to the formation of the earliest stars and galaxies at high 
redshifts, certainly $z>3.5$ (eg., Madau et al., 1996). 
Since these early objects would occur in regions destined to be 
incorporated into rich clusters (eg., BCFL, 
Governato et al., 1998) there seems little scope for weakening the
constraints by delaying the onset of galaxy formation.

In principle, observations of a narrow colour-magnitude relation at
higher redshift could constrain the galaxy formation epoch even more 
tightly (eg., Ellis et al., 1997). At $z\sim0.5$, the increase in the 
power of the test is modest, since the predicted rate of colour
evolution is still small. Moreover, the galaxy populations of
intermediate redshift clusters are considerably more heterogeneous than 
those of their present-day counter parts, and additional caution
must be exercised when deciding exactly which should be included
when the properties of the CMR are calculated. We therefore 
consider the properties of intermediate clusters separately in \S3.

\subsection{Continuous Star Formation}

\begin{figure}
\begin{center}
  \leavevmode
  \epsfxsize 1.0\hsize
  \epsffile{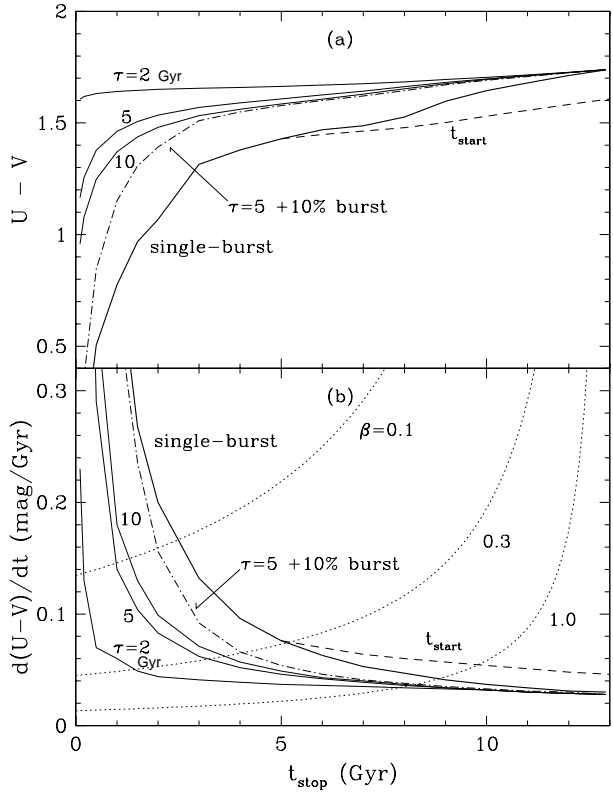}
\end{center}
\caption
{(a) $U-V$ colour and (b) rate of change of $U-V$ colour as a function of 
the `age' of the
composite stellar population. Solid lines are based on the models
of Kodama \& Arimoto, with abundance held fixed at the solar value.
Top thick line: a single burst model in which all stars are formed in
a short burst at $t_{stop}$. This model was used as the basis of 
BLE's diagram. Lower thin lines: stars are formed at a
decaying rate determined by the parameter $\tau$ until it is truncated at
$t_{stop}$. Three values are shown, $\tau = 10, 5, 2$~Gyr, from top to bottom.
Star formation is in each case initiated at $t_{start} = 13$~Gyr. 
Although the colours change very rapidly immediately after star formation 
finishes, the rate of change becomes small after several Gyr have elapsed,
and the movement of the galaxies in the colour-magnitude diagram
becomes negligible. The dot-dashed line shows effect of adding a 10\% burst 
of star formation to the $\tau=5\Gyr$ model at $t_{stop}$. The dashed line 
shows the rate of change of colour (with respect to $t_{start}$) when 
$t_{stop}$ is held fixed at 5~Gyr, while the epoch at which star formation 
begins ($t_{start}$, plotted on abscissa) is varied. Dotted lines in the
diagram illustrate the degree of coordination implied by the 
observed scatter in the colour-magnitude relation. The lines show
$\beta=0.1,0.3,1.0$ from top bottom. Further details are given in the text. 
}
\label{fig:ratefig}
\end{figure}

An alternative model for the star formation history of cluster
galaxies allows the star formation to take place over a protracted period
of time. This is conveniently modelled by an exponentially declining
star formation rate. Although the conventional view is that this star
formation occurs in a well-defined single event, this need not be the
case: the exponential decay is simply intended to characterise 
the change of the overall rate of star formation even if it is
split between several sub-units. We model the effect of the cluster
environment by truncating the star formation in the galaxy at the time
when it is accreted into the cluster environment. This is intended
to mimic the hostile effects of the dense environment, such as 
ram-pressure stripping by the dense intra-cluster medium 
(Gunn \& Gott, 1972, Kent, 1981, Couch et al., 1998) or high speed
galaxy-galaxy encounters (Moore et al., 1996). We have
avoided incorporating a more detailed model for these effects since
our over-riding aim is to produce a simple scheme that can be compared 
with the observations in order to investigate the coordination of the 
truncation event.

Our models assume that star formation can be parametrised by 
\begin{eqnarray*}
    SFR &\propto e^{t-t_{start}\over \tau} \qquad& t_{stop} < t < t_{start}\\
    	 &=0   & \hbox{otherwise}   
\end{eqnarray*}
Star formation starts at a time $t_{start}$, and declines at a rate
determined by $\tau$ until it is truncated at time $t_{stop}$. 
We fix the metal abundance of the system at the solar value.

Initially we consider the case in which star formation starts in all
galaxies at a similar high-redshift epoch and so set $t_{start}=13$~Gyr.
The rate of change of colour is shown as the thin solid line in Figure~2
for the values $\tau=2,5,10$~Gyr. As expected, the rate of change
of colour declines much more rapidly than in the case considered in 
the previous section. This is particularly true for the lowest $\tau$
models. Here very few stars are being formed at low look-back times.
When star formation is truncated, very little blue flux is produced
once the most massive stars have evolved off the main sequence. The 
presence of the remaining young stars has little effect
on the overall colour. Even in a galaxy with a long star formation
decay timescale, the rate of change of colour is lower
than the single-burst formation scenario discussed in the previous
section.

It is clear from this experiment, that the colour-magnitude relation
sets relatively weak constraints on the last epoch of star formation
in these systems. 
%
%
However, the approach of comparing the observed scatter with the
rate of colour change is a poor approximation if star formation occurs
over a prolonged period. Furthermore, the clipping procedure used to 
measure the observed scatter is not taken into account. A more
accurate method of assessing the scatter is to simulate the colour
distribution across the CMR and to then reproduce the observational
measurement.
%

\begin{figure}
\begin{center}
  \leavevmode
  \epsfxsize 1.0\hsize
  \epsffile{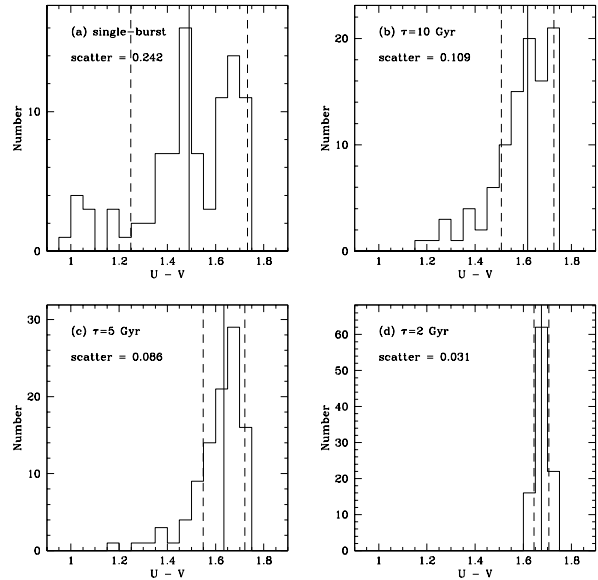}
\end{center}
\caption
{Simulated histograms of the distribution of galaxies across the 
colour-magnitude relation. (a)~Star formation occurs in a single burst.
The burst events are distributed uniformly between $t=0$ and 13~Gyr.
(b)~Star formation begins in all galaxies at $t=13$~Gyr, and declines at
a rate $\tau=10$~Gyr until it is truncated at $t_{stop}$. For different
galaxies, $t_{stop}$ is distributed uniformly between $t=0$ and 13~Gyr. 
(c)~as b, but $\tau=5$~Gyr. (d)~as b, but $\tau=2$~Gyr. In each case, the
vertical solid and dashed lines show the bi-weight average and scatter.
}
\label{fig:cmsimfig}
\end{figure}

The transverse cross-section of the model CMR is illustrated in 
Figure~\ref{fig:cmsimfig}. These histograms have been created by 
drawing 100 galaxies with random values of $t_{stop}$ in the range
0 to 13~Gyr. Each of the histograms shows the results for
different star formation models: (a)~all the stars form in a single
burst; (b)~star formation begins at 13~Gyr, and decays with e-folding
timescale $\tau=10$~Gyr until it is truncated at $t_{stop}$; (c)~and
(d) show similar models but with $\tau=5$ and 2~Gyr respectively.
The simulated galaxies have the same metal abundance and thus this diagram
neglects the effects of the slope of the CMR: in practice the scatter we 
derive from our models may be a slight underestimate due to the increased
brightness of the bluer galaxies. The effect is negligible, however, 
since the slope of the CMR is shallow relative to the changes in brightness.
For example, in the case of the $\tau=5\Gyr$ model, the increase in
dispersion due to this effect is 0.01~mag at most.

In order to make a quantitative comparison with observational data,
we use the bi-weight dispersion estimator (Beers et al., 1990).
This provides a measure of the scatter that is robust to deviations
from a Gaussian distribution. Its effect is to reduce the weight given
to the tail of extremely blue galaxies --- those in which star formation 
has very recently been truncated --- and to measure the characteristic 
dispersion of the population's colours. The procedure was repeated 1000 times 
in order to derive an estimate of the uncertainty in the bi-weight scatter.
The histograms in the figures show typical runs.

This analysis confirms the impression we derived from the differential 
rate of colour evolution shown in Figure~\ref{fig:ratefig}: forming
the entire stellar population in a single burst leads to a wide spread
in galaxy colours, but the scatter is much reduced if star formation
is spread over an extended period ($\tau=10$~Gyr). 
As expected, the colours become tightly bunched if the star formation is 
strongly weighted to the initial formation period since this was fixed
at 13~Gyr for all the galaxies in this simulation.

\begin{table*}
\caption{Scatter in simulated colour distributions. Star formation
in individual galaxies occurs between look-back times $t_{start}$
and $t_{stop}$ with a declining rate parameterised by $\tau$. 100 galaxies
are simulated with $t_{start}$ and $t_{stop}$ randomly distributed
as shown in the table. The columns give:
(1) the type of star formation model; (2) the lowest value used for the
$t_{stop}$ parameter; (3) the value or upper limit to the $t_{start}$
parameter; (4) The scatter of simulated colours calculated using the 
bi-weight statistic; (5) The 1$\sigma$ sample to sample variation in 
the scatter we measure for a sample of 100 galaxies.}
\vspace{0.5cm}
\begin{center}
\begin{tabular}{lrrrr}
\noalign{\medskip}
\hline\hline
\noalign{\smallskip}
Model&  $t_{stop,min}$&  $t_{start}$&   Bi-weight scatter\\
\noalign{\smallskip}
\hline
\noalign{\smallskip} 
single-burst&   0&     $= t_{stop}$&     0.242 $\pm$ 0.034\\
      &         5&            &          0.103 $\pm$ 0.004\\
      &        10&            &          0.029 $\pm$ 0.001\\
\noalign{\smallskip} 
$\tau=10$&      0&     fixed at 13&      0.109 $\pm$ 0.013\\
       &        5&            &          0.048 $\pm$ 0.002\\
       &       10&            &          0.014 $\pm$ 0.001\\
\noalign{\smallskip} 
$\tau=5$&       0&     fixed at 13&      0.086 $\pm$ 0.009\\
       &        5&            &          0.041 $\pm$ 0.002\\
       &       10&            &          0.013 $\pm$ 0.001\\
\noalign{\smallskip} 
$\tau=2$&       0&     fixed at 13&      0.031 $\pm$ 0.002\\
       &        5&            &          0.023 $\pm$ 0.001\\
       &       10&            &          0.011 $\pm$ 0.001\\
\noalign{\smallskip} 
$\tau=5$&  $0 - t_{start}$&   $0 - 13$&   0.316 $\pm$ 0.050\\  
	&  $0 - t_{start}$&   $5 - 13$&   0.138 $\pm$ 0.015\\  
	&  $5 - t_{start}$&   $5 - 13$&   0.084 $\pm$ 0.005\\  
\noalign{\smallskip}
\noalign{\hrule}
\noalign{\smallskip}
\end{tabular}
\end{center}
\end{table*}

\begin{figure}
\begin{center}
  \leavevmode
  \epsfxsize 1.0\hsize
  \epsffile{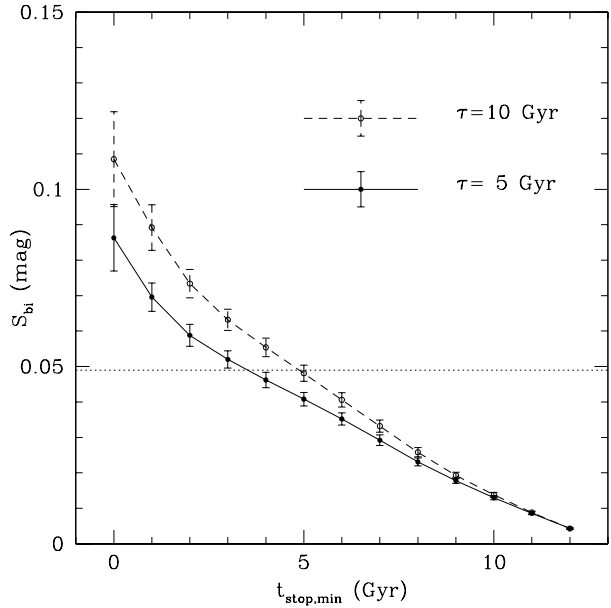}
\end{center}
\caption{The bi-weight scatter in the simulated scatter in the colour
magnitude relation as a function of $t_{stop,min}$, the look-back time 
to the epoch at which star formation in all galaxies. Two
models for the evolution of star formation are shown, $\tau=5\Gyr$
(solid line) and $\tau=10\Gyr$ (dashed line). Further details of
the star formation histories consider are explained in the text.
Error bars shown on the figure illustrate the 1$\sigma$ variation in 
the scatter measured for different Monte Carlo realisations of a sample
of 100 galaxies. A thin horizontal line illustrates the observed CMR scatter.} 
\label{fig:scatter}
\end{figure}

\begin{table}  
\caption{Observed bi-weight scatter for galaxies in the Coma cluster
from Terlevich et al., 1998, as a function of radius and morphological 
type. The sample is limited to galaxies with $V<17$ within a $13"$
diameter aperture. The values given have been corrected for the
contribution of measurement uncertainties (0.025 mag).}
\vspace{0.5cm}
\begin{center}
\begin{tabular}{lrrrrc}
\noalign{\medskip}
\hline\hline
\noalign{\smallskip}
    Region&  Morphology&  $\rm N_{gal}$&  Bi-weight scatter\\
\noalign{\smallskip}
\hline
\noalign{\smallskip}
 core ($r<20'$)&     E only& 	29& 	0.036\\
               &     E + S0& 	80& 	0.046\\
		&  all galaxies&   98&	0.049\\
      $r<50'$  &   all galaxies&  165&	0.061\\
\noalign{\smallskip}
\noalign{\hrule}
\noalign{\smallskip}
\end{tabular}
\end{center}
\label{tab:obstable}
\end{table}

A wider range of star formation histories is explored in Table~1. 
Here we show the effect of stopping all star formation at progressively 
earlier epochs (larger look-back times) by limiting the range of 
$t_{stop}$. As expected, the models in which star formation is truncated
at earlier times show lower scatter in colours. The importance of the
truncation obviously depends on the e-folding time of the model star
formation rate. For example, the $\tau=2$ Gyr model results in little 
scatter even if some galaxies are truncated as recently as the present day.
Figure~\ref{fig:scatter} shows how the scatter varies as a function
of the minimum value of $t_{stop}$ for star formation rates declining
as $\tau=5$ and 10~Gyr. For a given observational scatter, this can
be used to estimate the lowest look-back time at which star formation
can have occurred in a portion of the galaxies.
A model which considers the effect of varying both the start of star 
formation and its truncation is considered below.

\begin{figure}
\centerline{\psfig{figure=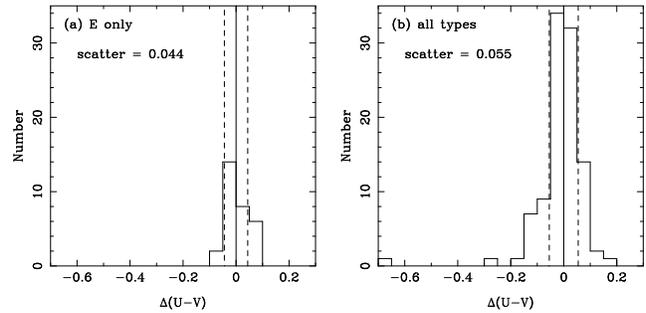,width=3.3in}}
\caption{The observed cross-section of the colour-magnitude relation
in the core of the Coma cluster. (a)~Bright elliptical galaxies in the
cluster core with morphological classification from Andreon et al., 1996.
(b)~The CMR for all bright galaxies regardless of morphological 
type. Galaxies brighter than $V<17$ and within the central $20'$
dia.\ of the cluster are plotted. The bi-weight scatter quoted here 
has not been corrected for observational uncertainties.
}
\label{fig:comascatfig}
\end{figure}

For comparison, Figure~\ref{fig:comascatfig} 
shows the histogram for the colour distribution
of galaxies in the core of the Coma cluster. This data is taken from
a wide-area $U-V$ Mosaic of the cluster (Terlevich et al.\ 
1998, summarised in Table~\ref{tab:obstable}), with colours calculated within 
an $8.7$~Kpc dia.\ aperture. Here we focus on the data taken from the core 
of the cluster, where accurate mophological types are available from
Andreon et al.\ (1996), although the full photometric data-set, extending 
to $50'$ from NGC~4874, has similar properties. Comparable results are 
obtained if we use data from BLE, although the galaxy selection is limited 
to early-types. 
By using Terlevich et al.'s data, it is not necessary to define 
a morphological classification system for the simulated galaxies since we
may consider colour properties of the galaxies independently
of their morphological appearance. In the core
of the cluster, early-type galaxies are dominant
and the measured scatter is not overly dependent on the method used to 
define the ridge-line of the CMR. 

The first panel of Figure~\ref{fig:comascatfig} shows the scatter in 
colour about the mean relation for 
elliptical galaxies only. For each galaxy, its 
deviation from the mean relation was calculated and then used to compile
the histogram shown in the figure. 
Selection of galaxies by morphological type does not accurately 
reflect the synthetic histograms that we have created, however. For example, 
galaxies with more recent star formation activity have predominantly 
disc morphologies. Because of the complete spatial coverage of Terlevich 
et al.'s data, we can also fit to the CMR without regard to galaxy 
morphology. This gives a similar result with just slightly larger 
scatter (Table~\ref{tab:obstable}). The difference is not surprising since 
most galaxies in the core of the cluster are early-types, and it is widely 
established (eg., BLE) that there is little difference in the properties 
of E and S0 galaxies in the cores of rich clusters.

Comparison with Table~1 shows that while allowing star formation to continue 
in some galaxies to the present-day produces too much scatter, truncating 
star formation before a look-back time of 5~Gyr reduces the scatter
to be acceptably small. In contrast to the single burst models that we
considered in \S2.1, strong coordination of the last epoch of star 
formation is not necessary in order to create a narrow colour-magnitude 
relation. Although the star formation rate decays slowly, the bulk of 
stars are still formed at large look-back times. A quantitative estimate of
the epoch at which the last stars were formed can be obtained from
Figure~\ref{fig:scatter}. The observational estimate of the scatter in the 
CMR given in Table~\ref{tab:obstable} (all galaxies, $r<20'$) of 0.049 
suggests that star formation may continue in a fraction of galaxies until a 
look-back time of 5~Gyr for $\tau=10\Gyr$, or 3~Gyr for $\tau=5\Gyr$.

These values allow considerably more star formation activity to have occurred
in the intermediate redshift universe than was suggested by consideration
of the single burst models: because
the complete galaxy population started to form stars at a similar
epoch, we have tended to ensure that the bulk of the stars have very similar
ages and colours. As a result, low scatter is measured when star formation 
is truncated early. To show the importance of the epoch at which star
formation is started, we fix $t_{stop}$ at 5~Gyr, and allow $t_{start}$ to 
vary. Dot-dashed lines in Figure~\ref{fig:ratefig} show 
the rate of colour change (with respect to $t_{start}$) as a function of 
$t_{start}$. As expected, these lines lie close to the burst formation model 
confirming that the colour-magnitude relation still places 
significant constraint on the epoch of the formation of the bulk of the 
stars (or its degree of coordination).
A model in which both $t_{start}$ and $t_{stop}$ are allowed to vary
cannot simply be represented in the Figure~\ref{fig:ratefig}. 
It is possible, however, to make Monte-Carlo simulations of this case by 
at first selecting $t_{start}$ with a uniform distribution and then 
selecting $t_{stop}$ to lie between this and some minimum value. 
As expected, these models produce larger variations in the CMR than when 
$t_{start}$ is held fixed. In this case even when no star formation occurs 
at look-back
times less than 5 Gyr, the variation in colours is still greater than 
the observed relation. The key result of this section is that the scatter 
of the C-M relation places a strong constraint on the formation epoch
of the bulk of the stars, but a relatively weak constraint on the
last epoch of star formation in the cluster galaxies. Thus the conclusion
reached by BLE that the bulk of star formation occurs at redshifts greater
than 1 (an average redshift of 2) appears robust; however, recent 
truncation of residual star formation is also compatible with the tight 
relation.

\section{Comparison with the Butcher-Oemler effect in Intermediate
    Redshift Clusters}

In this section we contrast the homogeneity of the stellar populations
of galaxies in nearby clusters with the properties of galaxies in 
clusters at intermediate redshift. Our aim is to examine the broad
compatibility of these observations. A detailed comparison of 
morphologically selected subsamples of galaxies will be made in a 
subsequent paper.

The observations of intermediate redshift clusters reveal significant
evolution in the galaxy populations of rich clusters. 
This is seen in two distinct ways. Firstly, through the changing
morphological content of clusters, originally observed as an increase
in the fraction of blue galaxies (Butcher \& Oemler, 1978), and now 
(post-HST) identified as an increase in the fraction of galaxies
with spiral and irregular morphology perhaps associated with a decline 
in the S0 galaxy content of distant clusters (Smail et al., 1997,
Dressler et al., 1997). 
We will refer to this as the morphological Butcher-Oemler effect.
Secondly, evolution is apparent in the fraction of galaxies with spectra 
showing abnormally strong Balmer absorption lines
(eg., Dressler \& Gunn, 1983, Couch \& Sharples, 1987, Couch et al., 1997,
Dressler et al., 1997). The most extreme H$\delta$ line strengths can only be 
reproduced if star formation is truncated after a strong burst of 
star formation (although the star burst may have a strongly skewed stellar
IMF). We will refer
to this phenomenon as the spectroscopic Butcher-Oemler effect. It is not 
clear whether the two versions of the Butcher-Oemler effect reflect the
same physical process (eg., Charlot \& Silk, 1994, Couch et al., 1997):
for example, the star burst fraction maybe driven by interactions with
gas rich galaxies while the general decline in the blue fraction might
result from ram pressure stripping.

At first sight, the evolution seen in the distant clusters appears to 
conflict with the homogeneity of the colour-magnitude relation in
systems nearby. Because galaxies cannot escape from these clusters,
star formation histories of the type discussed above must be typical of
many of the galaxies seen in our Coma cluster sample. Even
if these galaxies do not have regular E or S0 morphological types, the
data of Terlevich et al (Table~\ref{tab:obstable}) shows that similar scatter is obtained for the
present-day CMR even if morphological information is ignored.
The comparison can only be avoided if the galaxies responsible for
the Butcher-Oemler effect are systematically destroyed (for example
by `harassment', Moore et al., 1996) to form the diffuse intra-cluster 
light, or are confined to the outer-parts of present-day clusters.
We will assume that the first process is not efficient in what follows.
The second restriction is difficult to quantify because it is 
dependent on the efficiency with which these galaxies are mixed into
the central parts of the cluster. However, Allington-Smith et al.\ (1993)
concluded that there was no evidence for a radial gradient in the frequency of
starburst/post-starburst galaxies. Furthermore, Terlevich et al.'s data
for the Coma cluster extends out 2~Mpc (60\% of the 
cluster's virial radius) if the full area is considered, and 
incorporates the infalling NGC~4839 group in the South West of the cluster. 
This area is comparable 
with the region over which the Butcher-Oemler effect has been established 
at intermediate redshift (eg., Dressler et al., 1997, Couch et al., 1997, 
but see Abraham et al., 1996, van Dokkum et al., 1998 for studies with 
wider coverage). It therefore seems unlikely that the contrast in star 
formation histories can be explained by a simple radial gradient in
galaxy properties.
Finally, we note that the comparison requires us to make a further
assumption: that the clusters identified at intermediate redshift will,
by the present-day, evolve into systems with properties similar to those
of the Coma cluster. This comparison appears secure since 
the intermediate redshift clusters can only grow in richness, and the
tightness of the CMR appears a generic property of all
present-day rich clusters (BLE, Garilli et al., 1998).

\begin{table}
\caption{Bi-weight scatter in simulated present-day colour distributions
for a range of star formation histories compatible with observations
of the Butcher-Oemler effect. Model~1 assumes that all cluster galaxies
undergo truncation of their star formation at a random time between $t=0$ and 
13~Gyr. Model~2 mixes a 50\% population of these galaxies with a 50\% 
population of galaxies that cease star formation between 10 and 13~Gyr.
Model~3 is similar to Model~1, except that the distribution of truncation
times is taken from the infall rate given by Bower (1991). Model~4 mixes
Model~3 with a 40\% population of galaxies that cease star formation 
between 10 and 13~Gyr.
Further details of the models are described in the text. The quoted
uncertainties give the 1$\sigma$ variation between Monte Carlo realisations
of 100 galaxies.}
\vspace{0.5cm}
\begin{center}
\begin{tabular}{lrrrr}
\noalign{\medskip}
\hline\hline
\noalign{\smallskip}
Model&  $\tau=5$~Gyr&  $\tau=10$~Gyr\\
\noalign{\smallskip}
\hline
\noalign{\smallskip} 
 (1)&	0.122 $\pm$ 0.018 & 0.136 $\pm$ 0.021\\
 (2)&	0.052 $\pm$ 0.009 & 0.054 $\pm$ 0.010\\
 (3)&	0.075 $\pm$ 0.010 & 0.085 $\pm$ 0.012\\
 (4)&   0.052 $\pm$ 0.009 & 0.059 $\pm$ 0.010\\
\noalign{\smallskip}
\noalign{\hrule}
\noalign{\smallskip}
\end{tabular}
\end{center}
\label{tab:botab}
\end{table}

As we have shown, the scatter about the CMR constrains the 
formation of the bulk of the stars much more strongly than it limits 
the formation of the last stars. Is it possible then that the CMR 
scatter and the direct evidence for evolution are compatible?
To answer this question, we use the analysis of the star formation cycle
in distant clusters from Barger et al.\ (1996). They show that 
the relative numbers of star burst, post-star burst and red 
H$\delta$-strong galaxies can be described by the addition of a 10\%
(by mass) burst of star formation to an underlying spiral or 
elliptical-type (it is very difficult to estimate the star 
formation history of the galaxy prior to the star burst, 
Charlot \& Silk, 1994). After the burst, all star formation ceases.
A simple single cycle need not apply to all galaxies (for instance,
Barger et al.\ take no account of the relative magnitudes of the different
types of object). Nevertheless, several Gyr after star formation ceases, 
the strength of the star burst has a relatively small effect on the colour 
evolution  (Figure~\ref{fig:ratefig}) and so this simple 
parameterisation is adequate. Below, we assume that all galaxies pass
through this burst phase immediately after the truncation of the normal
mode of star formation.

We investigate the compatibility between the cluster population at $z=0.5$
and the scatter in the present-day population by developing the Monte-Carlo 
realisation technique introduced in \S2. We initially based the progenitor
galaxy population on the $\tau=5$~Gyr model so as to mimic the typical colours 
of present-day spiral galaxies. In order to
obtain the smallest possible CMR scatter, we assume that star formation
begins at around 13~Gyr in all systems. Truncation of star formation
at random times results in a roughly constant supply of galaxies
which then move redward onto the ridgeline of the CMR. This produces 
a constant decline in the fraction of spiral galaxies in the cluster.  
Observations of the `morphological' Butcher-Oemler effect (eg., Smail et al., 
1997) suggest a high rate of evolution of the spiral fraction: although a 
few percent at the present day, the fraction rises to $\sim 50\%$ at
$z=0.5$. A model in which star formation is truncated at a random
look-back time (cf., \S~2.2) produces a fraction star forming 
galaxies that varies linearly with look-back time. If all galaxies have 
an extended period of star formation the rate of evolution matches that
suggested by Smail et al (Model~1 in Table~\ref{tab:botab}), 
but tends to over estimate the rate of evolution suggested by the photometric
blue galaxy fraction. We reproduce the blue galaxy fraction by mixing 
the population of galaxies extended star formation histories with a 
population of intrinsically old systems. Our motivation for doing this
is that early-type galaxies may not represent a single homogeneous
population, and there may be several routes to the formation of
broadly similar present-day systems. We find that
a 50:50 mixture of systems that are intrinsically old and systems
with truncated star formation matches the typical slope of the `photometric'
Butcher-Oemler effect: at $z\sim0.5$, the cluster blue fraction is 
around 25\% (Model~2 in Table~\ref{tab:botab}).

\begin{figure}
\begin{center}
  \leavevmode
  \epsfxsize 1.0\hsize
  \epsffile{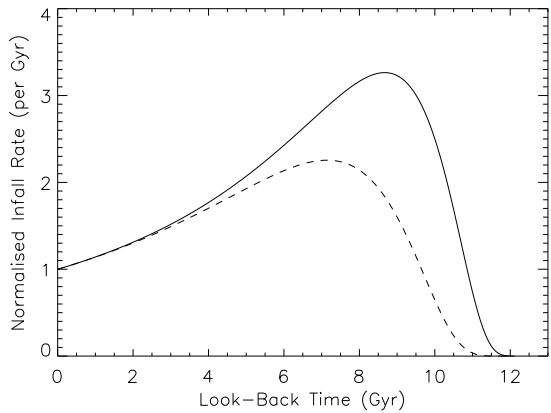}
\end{center}
\caption{The evolution with look-back time of the infall rate for a rich 
cluster. The extended Press-Schechter method has been used to calculate
the rate at which mass is in corporated into units more massive than a 
galaxy group (B91). These curves are used to suggest a parametric form 
for the rate at which star formation is truncated in the cluster galaxies.
Solid line: star formation is truncated on infall into corresponding
to a group of 5 or more galaxies ($M_s = M_*$, in the notation of
B91); dashed line, on infall into a group three times more massive
($M_s = 3 M_*$).}
\label{fig:infallfig}
\end{figure}

\begin{figure}
\begin{center}
  \leavevmode
  \epsfxsize 1.0\hsize
  \epsffile{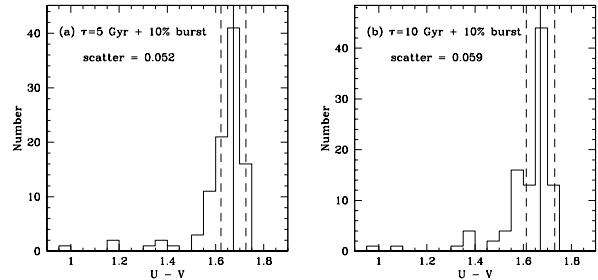}
\end{center}
\vspace*{-0.5\hsize}		
\caption{Simulated colour-magnitude histograms for a model of galaxy evolution
that is consistent with both the narrow scatter in the CMR
in local clusters and the observation of the Butcher-Oemler effect in 
high redshift systems. The model mixes a 50\% population of galaxies 
undergoing a random truncation of their star formation between $t=0$ and 
13~Gyr with a 50\% population of galaxies that cease star formation 
between 10 and 13~Gyr. The two panels distinguish assumption that
progenitors have star formation declining as (a) $\tau=5$ and (b) 10~Gyr.}
\label{fig:burstfig}
\end{figure}

Choice of a random uniform distribution of truncation times provides
a simple description of the process of galaxy infall and the termination
of star formation, but it is possible to use the analytic extensions of
the Press-Schechter theory (Bond et al., 1991, Bower, 1991 [B91])
to provide an improved the model for the rate at which galaxies
are accreted by the cluster. Using Eq.~18 of B91, and setting
$M_s=1.0$, gives the rate at which mass is accreted into objects more 
massive than galaxy groups. This formulae accounts for the total accretion
rate even if the final cluster consists of several fragments at earlier
times. We adopt the parameters $M'=100$, $n=-1.5$ from B91 in order
to provide a description of the growth of a rich cluster, and 
assume that the infall rate, $R_{\rm infall}(z)$, reflects the rate
at which star formation is truncated in the cluster galaxies (note that this 
approach differs from B91).  The evolution of the infall rate is plotted
in Figure~\ref{fig:infallfig}. The evolution of the rate of infall
is modest, increasing by a factor of 3 between the present and a 
look-back time of 9~Gyr and then rapidly declining at larger look-back times.
The function $R_{\rm infall}(z)$ is normalised and used to 
randomly select truncation times for Monte-Carlo simulation of the 
present-day CMR. This parameterisation assumes that star formation
is truncated when a galaxy is accreted into a system more massive than a 
group containing $\sim5$ galaxies.
We generate two models analogous to Models 1 \& 2.
Model~3 assumes that all galaxies are formed by truncation of exponentially
decaying star formation. This over estimates rate of increase in the 
blue galaxy fraction. In Model~4, we add a 40\% population of intrinsically
old galaxies so that the blue fraction increases more slowly with
redshift, reaching 25\% at $z=0.5$.

Table~\ref{tab:botab} shows the present-day scatter that we measure for each 
of the models. These models differ from the values given
in Table~1 because of the inclusion of the 10\% starburst at the final 
epoch. We first consider Model~1 in which the truncation rate is
constant and the actively forming galaxies in clusters rises
rapidly to a fraction of $50\%$ at $z\sim0.5$. This model produces
too many galaxies that are still moving towards the main colour-magnitude
relation. As a result, the predicted scatter exceeds that observed in
present-day clusters. Replacing the constant truncation rate by the 
rate derived from B91 (Model~3) slightly reduces both the blue fraction
at intermediate redshift (40\%) and predicted present-day CMR scatter,
although the latter is still too large to be compatible with the observations
of Terlevich et al. (1998). Altering the parameterisation of the truncation
rate by increasing, by a factor of three, the mass-scale at which star 
formation is switched off creates a distribution that is flatter and
thus tends to increase the predicted scatter. By contrast, adding a 
population of intrinsically old galaxies can dramatically reduce the 
predicted scatter. We add the maximum population that is allowed before
the blue galaxy fraction at $z=0.5$ drops below 25\%. With the 
constant truncation rate, adding a mixture of 50\% intrinsically
old galaxies reduces the rate at which galaxies are supplied to the 
CMR sufficiently that the predicted scatter provides a reasonable
match to that observed in the Coma cluster. The combination of the
high rate at which the colours of these galaxies redden, and the way
in which the mean relation is trimmed to remove outlying points weakens their 
influence on the bulk of the galaxies' properties. Similar results are 
obtained by adding a 40\% population of `old' galaxies to Model~3. 
The predicted present-day scatter is again 
within the observational limits. In these cases, evolution of the cluster
blue galaxy fraction from a few percent at the present day to $\sim 25$\%
at $z=0.5$ does not conflict with the observation of a narrow 
observed CMR in local clusters.

\section{Constraint on the Merger Histories of Cluster Galaxies}

In addition to placing a constraint on the star formation histories
of galaxies, the colour-magnitude relation can also be used
to set a limit on the amount of galaxy merging that occurs after the 
formation of the dominant component of the stellar population. 
We have shown in the 
previous section that the dominant component of the stellar population
was formed at relatively high redshifts. In this section, we address the 
question of whether these stars must already have been bound into a
single object, or whether they could have been formed in small units that 
were subsequently merged together. It is important here that we distinguish 
between mergers in which substantial star formation takes place and those 
that take place between systems consisting only of stars. The approach
we adopt aims to illuminate the general constraints that can be derived
regardless of the specific star formation model. As a result we concentrate
only on the second case, termed dissipationless merging by 
Bender et al.\ (1992a). 
The case in which mergers promote star formation cannot be considered 
in general terms and a specific model for star formation in galaxy mergers, 
and the conversion between hot and cold gas phases is required 
(eg., KC97, BCFL). 

The key ingredient to using the colour-magnitude as a constraint on 
galaxy merger histories is its slope. Even if the CMR is 
initially formed without scatter in a coeval galaxy population, disipationless 
mergers between systems of differing colour tend to average the 
colours of galaxies towards a single value. This reduces the slope of
the relation and increases scatter as galaxies undergo different numbers
of mergers between galaxies covering different ranges of colours. 
Thus, without knowing the initial slope of the relation, we can use the ratio
between the CMR scatter and its slope to place an upper-limit on the 
importance of dissipationless mergers.

We model this process by allocating galaxies to an initial colour
magnitude diagram. We assume that the initial relation is exact and that
there is no differential evolution of galaxy luminosities and colours. 
This is clearly an over simplification, but it distinctly separates the 
effects of the evolution of the stellar population (\S~2,3) from those 
due to the merging of galaxies. Under these conditions, the slope of the 
initial relation is the only parameter connecting the galaxy mass with
its colour. We emphasise that this scaling is arbitrary, and
that choice of a steeper initial slope simply rescales the final slope.
The key parameter that must be matched is the ratio between the scatter
about the average CMR and its slope. This ratio is independent
of the initial slope in the model.  

Experiment with random mergers between these objects shows that the 
initial relation is quickly weakened as galaxies are merged; however,
this approach does not accurately mimic the gravitational evolution 
of the galaxies' dark matter haloes. 
As KC97 have shown, 
there are significant correlations in galaxy merger histories meaning
that large galaxies form from systematically larger progenitors.
Therefore, we have refined our treatment to include the merger histories
as defined by the simulations of BCFL.  This has the added
advantage that it allows us to calibrate the merger process on to a 
(model dependent) redshift scale. We have included the results of
the equivalent random merger process as a means of determining the 
importance of the correlations present in the hierarchical model.

\subsection{Hierarchical clustering}

In order to generate galaxy merging histories, we use the formalism described 
in BCFL. This first generates a tree of merging haloes using 
the Monte-Carlo prescription outlined in Cole \& Lacey, 1993 (see also
Summerville \& Kolatt, 1998). In order to follow the merging history of 
galaxies, we must add an additional ingredient: the merging of galaxies 
within a common halo. This process is driven by dynamical friction, and we 
use the dynamical friction timescales as parameterised by Cole et al., 1994.
Galaxies that have dynamical friction timescales less than the
merger timescale of the halo are assumed to merge with the dominant
galaxy. Since the dynamical friction timescale is shorter for more
massive galaxies, this naturally generates a tendency for large galaxies
to form from mergers of massive galaxies. This approach is generic
to all hierarchical galaxy formation codes, and we use the code of 
BCFL to produce a list of galaxy fragments and their evolution
as a function of redshift. We emphasise that this is only used to 
produce the initial CMR and to select galaxies to be merged;
BCFL's parameterisation of the star formation process is ignored
in what follows.

It is important that we only select merger histories corresponding to 
galaxies that are found in rich clusters at the final time. We 
incorporate this requirement by only constructing merger trees for
objects bound into a dark matter halo with circular velocity greater than
1000~km/s at the final time. In order to reduce computational overhead,
also require that the final galaxy has absolute $V$-magnitude brighter
than $-19$ in BCFL's simulations. The final results are unaffected
by the cut off since we determine the slope and scatter over only the 
brightest 4 magnitudes of the simulated CMR (see \S4.3).

Once the galaxy merger tree has been constructed, our phenomenological 
approach diverges from the prescriptive modelling of BCFL. We 
start by selecting an initial epoch ($z_{form}$). The galaxy fragments 
present in the tree $z=z_{form}$ are given a magnitude scaled to the 
log of their total baryonic mass, and a colour according to an initial 
`perfect' CMR (ie., $\hbox{colour} = \hbox{slope}\times 
\hbox{magnitude} + \hbox{offset}$). 
As we have already emphasised, the `colour' we calculate is purely 
illustrative and serves only to show the effect of mixing the stellar
populations. We make no attempt to justify the initial zero-point or slope
that is applied. The evolution can then be followed by simply combining 
the magnitudes and colours of the fragments as they merge to form the 
final galaxies at $z=0$. At some points in the tree the proto-galaxies merge
with `new' fragments (fragments with progenitors that are below the mass
resolution of the merger code); in order to be consistent with our 
phenomenological model, we do not incorporate these objects in to 
the merger scheme. The treatment of these objects is only important
if the CMR is set at high redshifts $z>2.5$: at later times the
fragments are very much larger than the mass resolution limit.
Each tree leads to the formation of a single galaxy at the present epoch. 
In order to match the numbers of galaxies in our observational sample the 
process is repeated in order to generate a realisation of the CMR.

\begin{figure*}
\centerline{\psfig{figure=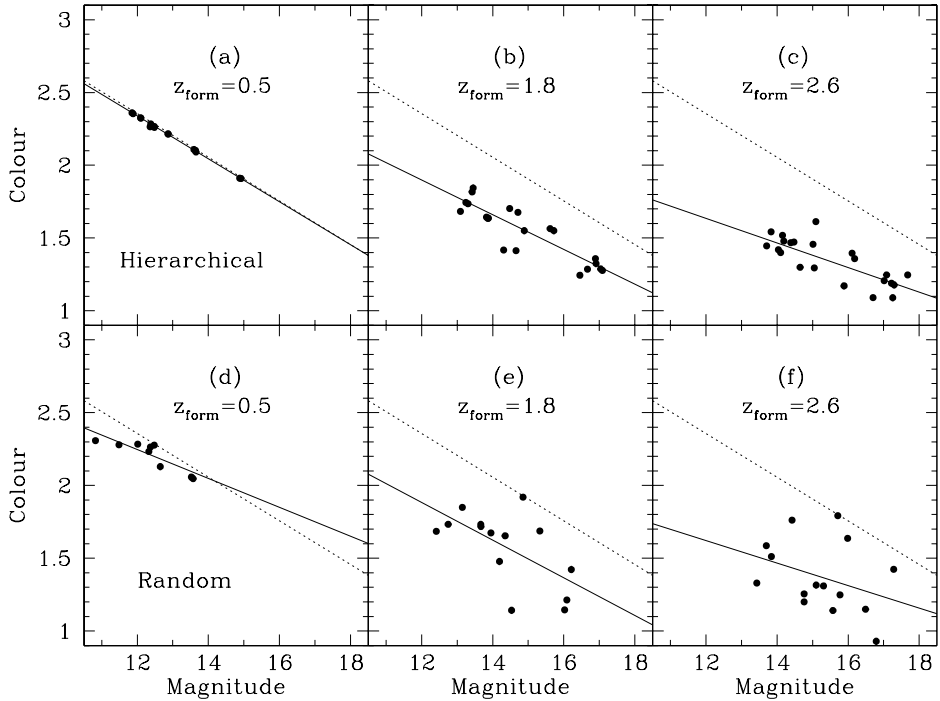,height=4in,angle=270}}
\caption
{Example Monte-Carlo realisations of the present-day (pseudo) colour-magnitude 
relations produced by the hierarchical clustering model (top row) and the 
random merging model (bottom row). Details of the procedures used to generate
these relations are given in the text. The solid line 
shows the mean CMR fitted by minimising the bi-weight scatter of 
the residuals, while the dashed line shows the initial relation. 
The three columns correspond to 
different redshifts at which the initial model CMR is implemented 
$z_{form}=0.5 , 1.8 , 2.6$  respectively. As can be seen the slope of
the relation weakens and the scatter increases as the degree of merging
between the initial epochs and the present-day increases. The effect is
more marked for a random merging process than for the hierarchical merger
model.}
\label{fig:models}
\end{figure*}

Panels a, b \& c of Figure \ref{fig:models} show examples of the present-day 
CMRs that we derive by this approach. If the relation 
is established at relatively low redshift, the rate of merging is low 
enough that little evolution of the initial slope occurs.
By contrast, a relation that is established at higher redshift is 
considerably weakened due to the large number of mergers that have taken 
place. Despite this, a discernible CMR exists at the present-day
even if the initial relation is established at $z>2.5$. We fitted the CMR
of the brightest four magnitudes of the relation by minimising the 
bi-weight scatter. This is shown by a solid line in each of the panels.
The degree of robustness is surprising. As we show below it is due
to the slow rate of mass growth in the hierarchical merger tree at late
times.

\subsection{Random mergers}

In order to test the importance of the merger correlations in the 
hierarchical model, we repeated the Monte Carlo process using
randomised merger trees. In order to be able to compare our results 
directly, we have based the simulation on the same galaxy merger trees 
to imprint the initial CMR at $z=z_{form}$. However, rather than 
merging the fragments as ordered by the hierarchical tree, 
random fragment pairs (possibly from different trees) are selected at each 
branch of the original tree. Fragments without progenitors in the 
hierarchical tree also have no progenitor in the random tree.  
This approach preserves both the initial mass distribution and number
weighted merger rates. 

As can be seen from Panels d, e \&~f of Figure~\ref{fig:models}, the effect 
of the random mergers is to more rapidly reduce the slope of the initial tree
and to rapidly increase the scatter of the CMR. Although a 
best fit line can be defined for these models, the slope varies considerably
between different realisations. The data used below have been averaged 
over many realisations of the random merger trees.

\subsection{Comparison with Observational Data}

We use the scatter in the present day CMR to constrain its
formation history, however the scatter that we measure in the simulated
CMRs can be made arbitrarily small by reducing the slope of the
relation. For example, initially allocating all galaxies the same colour
will (assuming a coeval population) lead to a final CMR with
zero scatter but also zero slope. The key requirement is therefore 
that the merging process is able to maintain the ratio between the 
scatter and slope ($R$) at or below the observed ratio. This is independent of 
our initial CMR, and allows us to compare both the models to each
other and to real data.

\begin{figure}
\centerline{\psfig{figure=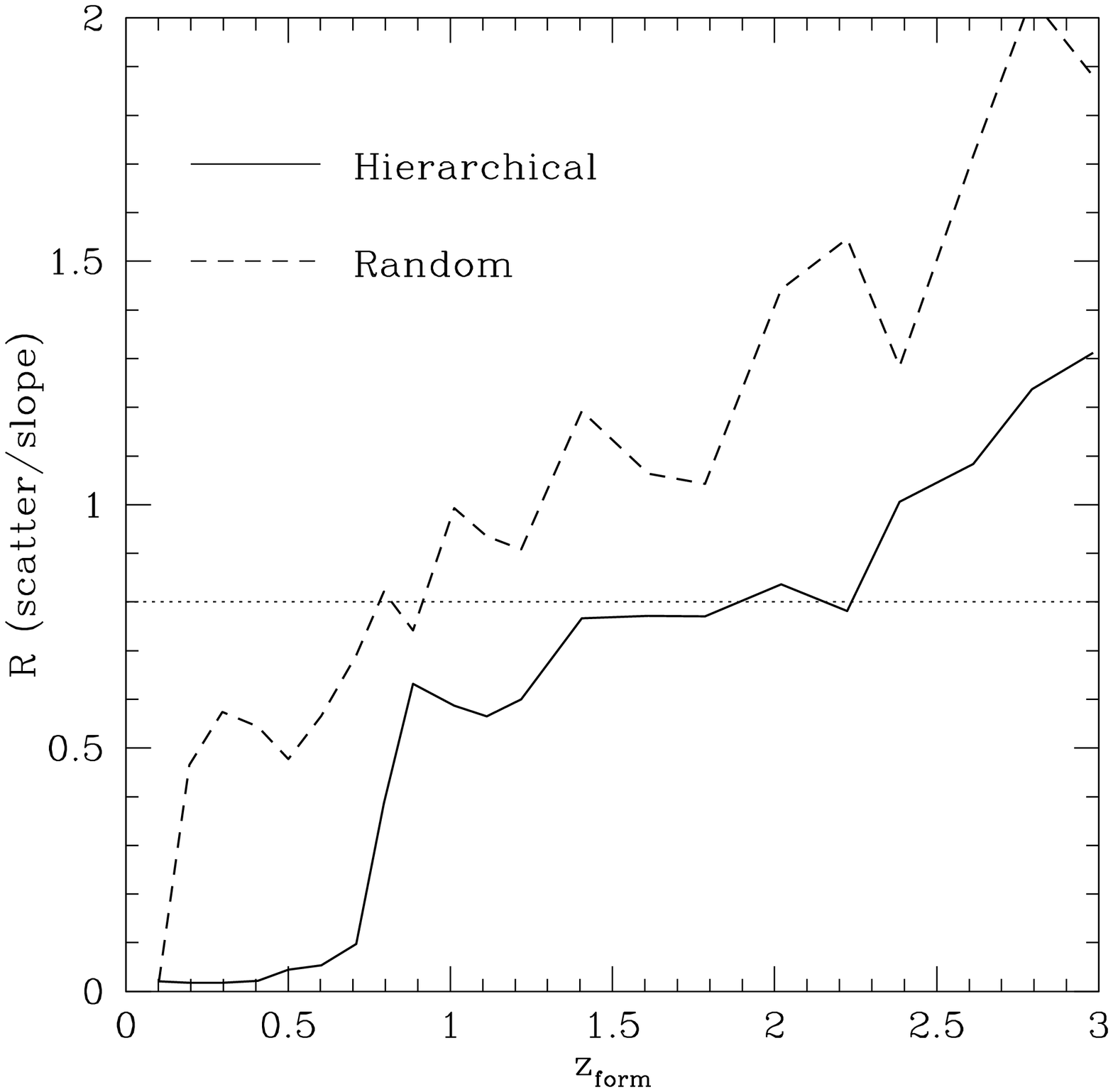,height=3in}}
\caption
{The effect of mergers in a hierarchical clustering scenario. The solid
line shows the how the ratio between present-day CMR scatter
and slope ($R$) depends on the redshift $z_{form}$ at which test galaxies 
are assigned colours according to an exact correlation 
between colour and magnitude at redshift. The effect of mergers in an 
equivalent random merging model is shown by the dashed line.
}
\label{fig:mergfig}
\end{figure}

Figure~\ref{fig:mergfig} summarises how the ratio $R$ varies with $z_{form}$.  
In both random and hierarchical models $R$ increases
monotonically, however the hierarchical clustering model has a much
slower increase of $R$ with $z_{form}$. 
We compare these results with our data from the Coma cluster. The
observed $U-V$ scatter is 0.049 (all galaxies, $r<20'$, Table~2) and
the observed CMR slope, measured using $U-V$ colours in a metric ($13''$) 
aperture and total $V$ magnitudes, is $-0.087$ (BLE).
This gives a ratio of 0.6. It is important to use the slope referenced to a 
total magnitude rather than an aperture measurement so that this quantity is
conserved during the mergers as in the model calculation. There is, however,
a further correction to this slope. The colours that available for Coma
are measured in a metric aperture that takes no account of the size of 
the galaxy. Thus larger galaxies have their colours measured within a 
smaller relative diameter. This will naturally tend to introduce a slope
into the CMR since galaxies become bluer at large radii.
To estimate the size of this effect, we use the elliptical galaxy
colour gradients measured by Peletier et al.\ (1990). Their mean colour 
gradient is measured in
$U-R$ and $U-B$, but it can be converted to a gradient in $U-V$ using
the late type stellar colours from Gunn \& Stryker, 1983. We find
$\Delta(\mu_U - \mu_V) / \Delta\log r = -0.178 \, \hbox{mag}/\hbox{dex}$. 
This can be converted to a change in aperture colour as a function of 
radius by integrating over the de Vaucouleurs $r^{1/4}$ profile, giving
a change in colour of $-0.134 \, \hbox{mag}/\hbox{dex}$. In order to 
assess the effect of this colour gradient, it is simplest to compare
with the observed slope of the colour-magnitude relation plotted in 
$U-V$ vs.\ $D_V$ coordinates ($D_V$ measures the size of the galaxy within
which the mean surface brightness is 19.80 mag arcsec$^{-2}$, Lucey et al.,
1991). The observed slope of in this parameter
space is $0.46 \, \hbox{mag}/\hbox{dex}$ (BLE), which should be compared
with the effect expected due to colour gradients of $0.134$
(the aperture used to measure the colour remains fixed and therefore the 
relative size of the aperture is inversely proportional to $D_V$). 
Thus the colour gradient effect accounts for about 30\% of the CMR slope. 
Including this correction raises the ratio of observed scatter to slope
to 0.8.

We can compare this limit with the simulated models. For the random model,
the ratio $R$ is already too large for $z_{form}\geq 0.9$. For the 
hierarchical model, $R$ increases less rapidly, so that the 
constraint on $z_{form}$ is correspondingly weakened, and the observed
limit on $R$ only conflicts with the model strongly for
$z_{form}\geq 2$ (although there is little room for additional 
sources scatter once $z_{form}\geq 1.4$). This limit is rougly consistent 
with the formation of the bulk of the stellar population at large
look-back times as discussed in \S2. Quoting these results in terms of 
redshift is, however, a little unsatisfactory
as the redshift evolution of the objects mass is tied to the specific
model for dynamical friction and the merger timescale adopted by 
BCFL in generating the galaxy merger tree. A more
useful classification is the factor by which the masses of the galaxies
have increased over these epochs. We use the ratio of the mass-weighted 
mean mass (MWMM) between the final CMR and the epoch at which the CMR is set 
as a measure of the amount of mass growth. This is analogous
to $M_*$ of the Press-Schechter mass function, and provides a measure
of the `typical' stellar mass of the model galaxy population. By assuming a
particular mass-to-light ratio for the stellar population, the MWMM
can be converted into an equivalent luminosity, however, this is not required
for the comparsion we wish to make. The change in the MWMM ratio provides a 
simple measure of the mass a typical galaxy has gained in the 
merger process. In the heirarchical tree, this is not equivalent 
to the number of mergers: as we will see, the mergers at late times 
(low redshifts) are dominanted by the accretion of small objects 
that make little change to the MWMM. 

\begin{figure}
\centerline{\psfig{figure=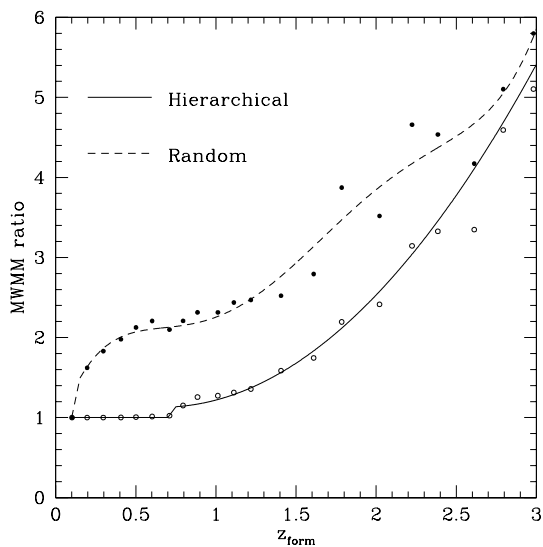,height=3in}}
\caption
{The evolution of the MWMM, expressed relative to the $z=0$ value, as a 
function the formation redshift $z_{form}$
a heirarchical galaxy model (solid line) and a model with random
galaxy mergers (dashed line). Details of the models are given in 
the text.} 
\label{fig:lwmlevol}
\end{figure}

\begin{figure}
\centerline{\psfig{figure=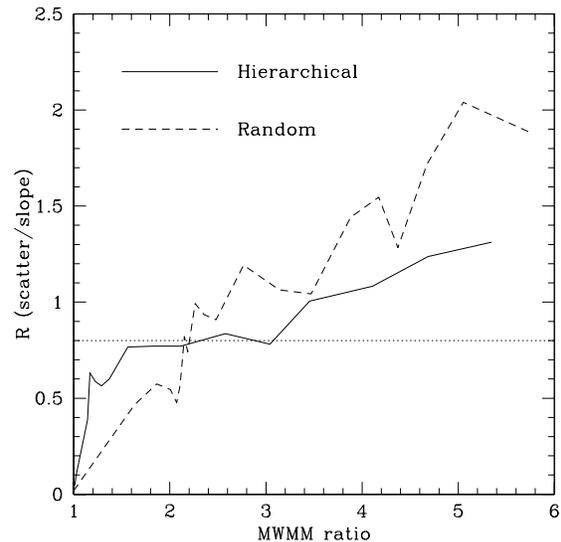,height=3in}}
\caption
{The evolution of the ratio between CMR scatter and slope ($R$) for 
a heirarchical galaxy model (solid line) and a model with random
galaxy mergers (dashed line). Details of the models are given in 
the text. The observed value of $R$ is shown as a thin horizontal line.} 
\label{fig:growfig}
\end{figure}

Figure~\ref{fig:lwmlevol} shows how the MWMM evolves as a function of
redshift for the hierarchical and random models. We first
consider the heirarchical tree. Initially, the MWMM ratio evolves
only slowly: at these redshifts most mergers involve the accretion of a 
small companion by a large galaxy. By $z_{form}=1$, the characteristic
mass has grown by only 20\%, and the rate of growth remains low
until $z_{form}>2$. By contrast, the evolution of the MWMM of the random model
is more sensitive to the choice of $z_{form}$. This a result of the 
way in which the random model is constructed from the same galaxy
fragments as present in the heirarchical model at the initial epoch.
It is important to note that Figure~\ref{fig:lwmlevol} does not show
the evolution of the MWMM for a single realisation, but rather the effect
of different starting points for the random merger tree. 
At low redshifts, the initial mass distribution is dominated by high
mass objects: in the random model merging of these components has high
probability, and there is considerable evolution of the characteristic 
mass. At higher values of $z_{form}$, a greater fraction of the initial
mass is dominated by the smaller mass objects, and the behaviour becomes
more similar to that seen in the heirarchical tree. To quantify the mass
growth factor, both datasets have been interpolated by a smooth relation 
shown by the thin lines in Figure~\ref{fig:lwmlevol}. We use this convertion
to compare the evolution of the CMR as a function of mass growth. 
This is shown in Figure~\ref{fig:growfig}. Despite the large differences
in the rate of mass growth, the evolution of the slope to scatter ratio, 
$R$, is similar for both the random and heirarchical merger trees. 
Both curves show steps in the $R$ that result from the non-linear behaviour
of the bi-weight scatter estimator. Although the overall impression
is one of similarity, the correlations inherent in the heirarchical merging
tree result in the scatter from this model exceeding the random model at 
low mass growth factors and under shooting it at high values. Nevertheless,
when compared to the observed value of $R$, both models set similar 
limits on the mass growth factor. Although the CMR
can be imprinted at higher redshift in the heirarchical tree, this is
compensated by the slower growth of mass in this model. 

It is interesting to compare these results with KC97.
While our approach differs from their global modelling of the galaxy
formation process, we find similar conclusions for the factor by which
galaxy mass has increased since the formation of the bulk of the stars.
The key point is that this factor is sufficiently small (between 2 and~3) 
that the CMR is adequately able to retain a memory of its initial slope. The
particular model by which galaxies are selected for merging seems to have 
only a weak effect on the mass growth factor, although it has significant
implications for rate of redshift evolution. 

The existence of a strong colour-magnitude correlation
rules out the possibility that typical cluster galaxies have grown in mass 
by a large factor since the formation of the bulk of their stars.
Furthermore, the factor of 2-3 that we estimate probably
represents an upper limit to the role of dissipationless 
mergers in the formation of (the majority of) present-day luminous 
cluster galaxies. 
For example, it is extremely unlikely that the initial CMR will be
completely perfect. Furthermore, we have not allowed for the
differential evolution of galaxy colours as discussed in \S2. These effects
will tend to make the limit on the degree of merging still more
stringent. The constraint can only be weakened if the mergers are not 
purely disipationless and are associated with significant star formation 
and consequent metal enrichment. This then conflicts with our initial 
parameterisation of the CMR. 
We discuss the role of these effects further, and examine
possible explanations for the origin of the CMR in the following 
section.

\section{Discussion}

At first sight, the early formation of the bulk of stars in cluster
galaxies presents a problem for the heirarchical scenarios of galaxy
formation. However, while star formation generally occurs late in these
models, clusters of galaxies are special parts of the universe in which
structure formation is more advanced. Thus the formation of the bulk
of stars at redshifts greater than unity does not present a particular 
problem since our inferences apply to only rich clusters. The difference 
between the formation of average structure and that leading to present-day
rich clusters can be seen directly from the extended Press-Schechter
formalism described by Bower (1991) and Bond et al.\ (1991). This allows
us to determine, as a function of present-day halo mass, the epoch at 
which 30\% of the final mass has become bound into objects of a certain 
scale size. For example, Moore et al.'s (1990) analysis of the CfA redshift 
survey suggests that the present-day mass distribution has a characteristic 
mass scale ($M_*$) equivalent to a group containing 5 $L_*$ galaxies and an 
effective power spectrum index, $n=-1.5$. With these parameters, 30\% of the 
universe becomes bound into galactic size objects at $z \approx 0.5$.
This is a very late epoch. By contrast, the matter that is destined
to be incorporated into present-day rich clusters ($M \sim 10 M_*$) reaches
the same threshold at $z\approx 1.4$. This simple calculation is reinforced
by the more detailed calculations that can be made using full galaxy
formation codes (eg., Kauffmann, 1996).

It is quite natural therefore that although the majority of stars are
formed at relatively recent epochs, the stellar populations of galaxies
found in rich clusters are biased to form at
high redshifts. This is in broad agreement with the small scatter we
observe in the colours. From this argument alone, however, it is not clear
whether these stars are formed in sub-galactic fragments and only later 
combined into the cluster galaxies that we observe. 
We have addressed this issue by investigating the effect of mergers on the 
slope and scatter of the CMR. Extensive dissipationless merging 
flattens the CMR and introduces additional scatter. This happens
rapidly if the merger process selects galaxy fragments at random.
Random merging is not a good description of the process expected in a 
heirarchical model, however, since most mergers involve
the galaxy at the centre of the halo and its satellites. Encounters between
the satellites themselves have high relative velocity leading to 
`harassment' (Moore et al., 1996) rather than merging.
The dynamical friction timescale is strongly sensitive 
to the ratio of the mass of the satellite to that of the 
halo as a whole (eg., Binney \& Tremaine, 1987). As a result, galaxy 
mergers of the most massive galaxies occur shortly after the formation
of a new halo, while galaxies of lower mass tend to remain
in orbit about the dominant object for a number of dynamical times. This 
introduces strong correlations into the merger tree so that mergers at
low redshifts tend to involve lower mass satellites. These contribute 
little to the mass of the brighter galaxies and there is little
flattening of the CMR. Because of these effects, it may 
be possible to construct a coherent model in which the star formation 
occurs at relatively early epochs and yet
the growth of galaxies by mergers does not overly weaken the initial CMR. 
One approach is to treat separately the constraints on the degree of 
mass growth and on the last epoch of star formation. Alternatively,
the two limits may be combined using the mass growth factor given by the
heirarchical merger tree.
For example, consider a model in which the star formation occurs 
over extended periods and is truncated randomly so that
star formation ceases in all systems before $z\sim1$ 
(ie., $t_{stop,min} \sim 8.5\Gyr$ in the terminology of \S2.2).
If the CMR has negigable scatter at $t_{stop,min}$, we can estimate the 
combined effects of the differential reddening and merging. 
Figure~\ref{fig:mergfig} suggests that
merging will result in a present-day $R$ value of 0.6, corresponding
to a scatter of 0.04~mag about the observed slope. Differential colour
evolution contributes a scatter of 0.02~mag (Figure~\ref{fig:scatter}). 
Assuming that these contributions can be added in quadrature, the model
would be well matched to the observed scatter. However, this estimate is
clearly conservative.
If the star formation rate declines as $\tau=10 \Gyr$, the $V$-band luminosity
weighted age of the stellar population is 10.5~Gyr (look-back time, $z=2$) 
in even the youngest galaxies. As a result, our estimate of the contribution 
to the scatter from galaxy merging has probably been 
underestimated. If the merger contribution is calculated from a look-back 
time of 10.5~Gyr, the combined scatter exceeds the observed value.  
Unfortunately, it is not possible to perform this calculation more rigorously 
without a detailed model connecting star formation, merging and chemical 
evolution.

A model in which all star formation ceases completely at early epochs
in cluster galaxies cannot, in any case, be reconciled with the increasing 
fraction of blue galaxies seen in intermediate redshift clusters.
This is a fundamental observational problem: if we accept
that the small scatter seen in the Coma cluster is representative of local 
clusters (cf., BLE, Garilli et al., 1996), and that
we have observed a sufficient range of cluster richnesses at high redshift
to encompass its likely progenitors (cf., Stanford et al., 1997, Smail
et al., 1997), then there must be a solution to this paradox. We would 
like to avoid extreme solutions that side-step the problem: for example, 
the blue colours of the intermediate redshift galaxies might arise from 
a top heavy IMF, or the blue light might be concentrated into stellar 
discs which are subsequently stripped in the cluster environment 
(Moore et al., 1996). Our modelling of the process by which
the scatter is measured shows that such conclusions are avoidable.
A model in which star formation in the cluster population is truncated
uniformly over the available cosmic time indeed produces an unacceptably
large scatter in the present-day CMR. However, mixing this
population with an equal population of older galaxies results in
a predicted CMR scatter that is compatible with the observed values and
accounts for the galaxy populations observed in intermediate redshift
clusters.

So far, we have considered the scatter across the CMR without discussing 
the origin of its slope. In models in which
star formation is completed over relatively short timescales, the
slope of the CMR has a natural explanation (eg., Arimoto \& Yoshii,
1987). Higher mass systems are better able to contain the star forming 
gas against the winds generated by the on-going super-novae. They thus 
reach a higher mean metal abundance before the energy released
by the supernovae is able to drive the gas out of the galaxy and
halt the formation of further stars. If we now propose that star
formation in early-type galaxies occurs over an extended period of
time, the connection between galaxy mass and the efficiency with
which gas is converted into stars (or equivalently the metal abundance
of the system) is much less evident. It is reasonable to ask whether
there should be any correlation between stellar mass and colour at
all.  There are two alternatives: the star formation process must
either be self-regulated by dynamic interchange of the gas expelled by 
supernova driven winds and the cooling gas from a halo reservoir (White \& 
Frenk, 1991, Kauffmann et al., 1993, Cole et al., 1994), or by episodic 
bursts of star formation driven by galaxy mergers (Larson \& Tinsley, 1974,
Balland et al., 1997). 
Both of these models are capable of explaining the slope of the
correlation, but the origin is very different in each. By itself,
a closed box model of chemical evolution tends asymptotically
to a metal abundance (the `yield', eg., Tinsley 1980) that depends only
on the initial mass function of the stars formed. However, if (1)~star
formation is allowed only to occur in discrete bursts when galactic
units merge and  (2)~the initial gaseous building blocks are all
of similar size, the rate at which a system progresses towards the
asymptotic abundance is linked to the number of mergers and hence
to the total mass. Nevertheless, the model's success is dependent on
the ad hoc assumption of a limited size range for the proto-galactic
units (otherwise galaxies of the same final mass may be built
from widely differing numbers of mergers) and the careful balancing of
the rate at which gas is consumed in mergers so that only the most
massive galaxies approach the yield. 

By contrast, a model in which the outflow of enriched gas is balanced
by inflow of less enriched material naturally produces a relationship
between galaxy mass and abundance since the rate at which the enriched
material is driven from the galaxy depends on its binding energy.
The result is an effective yield that varies with galaxy mass. In
low mass systems, most of the metals produced in supernovae escape
from the galaxy; those that remain are diluted by newly accreted 
material (White \& Frenk 1991). In high mass systems, most of 
the metals are retained in the galaxy and the inflowing matterial
is quickly enriched. The result is a strong correlation between 
mass and metal abundance with a slope determined by the relative
ease with which supernovae ejecta escape the galaxy. Such feedback
is required in heirarchical models in order to flatten the faint
end of the galaxy luminosity function (eg., White \& Frenk, 1991, 
Cole et al., 1993, BCFL). 
A consequence of this model is that the correlation should not be restricted
to early-type galaxies in clusters, but should also be observable
in general field spirals. The existence of such a correlation is
still controversial for the stellar disc (eg., Tully et al., 1982, 
Mobasher et al., 1986, Peletier \& de Grijs, 1998), although a 
correlation between HII region metal abundance and total brightness
is well established (Roberts \& Haynes, 1994, Zaritsky et al., 1994). 
A further consequence of this
model is that it is reasonable to treat separately the effects of passive 
aging of stellar populations and their metal abundances. 
The metal abundance of the stars in a galaxy is set by a 
combination of the yield of the stellar population, the
outflow of enriched material and the inflow of unenriched gass. These
combine to set an effective yield that is a function of the mass of the
galaxy's dark matter halo. Unlike the scenario of Arimoto \& Yoshii, 
the metal abundance does not depend critically on the duration for which
star formation continues; thus two White \& Frenk-type galaxies that
continue their star formation for different lengths of time 
will both have similar metallicities. Importantly, the longer timescale of
star formation in one galaxy cannot compensate for its bluer colours 
(due to younger stellar population age) through a higher metal abundance. 
In this model at least, there is no possibility of a conspiracy between metal
abundance and age that conceals large variations in star formation history.

\section{Conclusions}

The narrowness of the colour--magnitude relation in cluster galaxies
imposes a definite degree of homogeneity on their stellar populations.
BLE applied this argument in the context of single burst models
to conclude that the majority of star formation occured at look-back
times greater than $\sim 10\Gyr$. In this paper, we have explored a greater
variety of possible star formation histories. We find that the 
bulk of stars must still have old ages but that star formation continuing
until relatively recent epochs can be accommodated within the observed
CMR scatter in local clusters. This conclusion agrees with that
of van Dokkum et al., 1998, who used analytical models for the evolution
of galaxies luminosities to explore a wider range of
possible star formation histories. As a result, the rising fraction of 
blue galaxies in clusters out to $z\sim0.5$, as observed in the 
Butcher-Oemler effect, does not conflict with the homogeneity 
of the majority of cluster galaxies. Importantly, these constraints
on the star formation histories of cluster galaxies can be naturally
reconciled with hierarchical models of galaxy formation since 
the cluster galaxy population is assembled at substantially higher redshifts 
than galaxies drawn at random from the universe as a whole. 
 
The CMR can also be used to limit the degree of
merging that can occur between galaxies after the stellar population has 
been formed. Random mergers quickly flatten the CMR
leading to a high ratio between the scatter and slope. With random merging,
this point is reached quickly, so that the bulk
of stars in these galaxies must be formed at redshifts less than 1. 
However, merging between galaxies is not well modelled by a random process
since the dynamical friction time scale is strongly mass dependent.
We have incorporated the effect of correlated mergers using galaxy 
merger trees taken from BCFL. This substantially lengthens
the survival time of a recognisable CMR. Even at $z=2$, the ratio of the 
scatter to slope is kept small, lying just within acceptable bounds.
As a result, it is possible to construct a model for the formation of
cluster galaxies in which the bulk of the stars are formed at high
redshifts (resulting in little differential colour evolution at the 
present-day) and yet undergo sufficiently little merging that the 
colour-magnitude relation is still well-defined at the present-day.
Nevertheless, the factor by which the mass of a typical galaxy grows
between the formation of the bulk of the stars and its present-day mass
is small, less than a factor of 2--3 regardless of the merger tree considered.

This conclusion agrees with the study of KC97, 
who found that in their specific galaxy formation model the CMR
of early-type galaxies was well defined and showed scatter 
comparable to the observational measurements. Our approach
has been complementary: a tightly defined colour-magnitude relation at 
low redshift is not a property related to a single specific model, but 
will be met by any model in which most star formation occurs at early
epochs ($z>1$) in cluster galaxies and in which subsequent
mass growth is not too large. We have avoided distinguishing galaxies on 
the basis of their morphology by using robust algorithm to measure the 
scatter of the mean relation with all morphological types present. This
results in a conclusion that is as model independent as possible.

\section*{Acknowledgements}

We are indebted to Carlton Baugh, Shaun Cole, Carlos Frenk and Cedric
Lacey for allowing us to use the output from their heirarchical galaxy 
merging trees in order to investigate the effect of galaxy mergers on 
the evolution of a pre-existing colour magnitude relation, and to
Nelson Caldwell for allowing us to use the results of our Coma cluster
collaboration prior to publication. We would also
like to thank the referee, Guinevere Kauffmann, for her constructive
comments, and Alfonso Aragon-Salamanca and Richard Ellis for helpful
discussions. 

T.K. thanks JSPS Postdoctoral Fellowships for Research Abroad for
financial support. RGB acknowledges the support of a PPARC Rolling Grant
``Extragalactic Astronomy and Cosmology at Durham''. This project has made 
exensive use of Starlink computing facilities.

\end{document}